\documentclass[preprintnumbers,amsmath,amssymb,aps,eqsecnum]{revtex4}
\usepackage[dvips]{graphics,color}
\usepackage{latexsym}
\usepackage{graphicx}

\def\be{\begin{equation}}
\def\ee{\end{equation}}
\def\bea{\begin{eqnarray}}
\def\eea{\end{eqnarray}}
\begin{document}

\title{Non-linear interactions in a cosmological background in the DGP braneworld}

\author{Kazuya Koyama$^1$ and Fabio P Silva$^1$}

\address{
 $^1$Institute of Cosmology \& Gravitation,
University of Portsmouth,
Portsmouth~PO1~2EG, United Kingdom\\}

\begin{abstract}
We study quasi-static perturbations in a cosmological 
background in the Dvali-Gabadadze-Porrati (DGP) braneworld model. 
We identify the Vainshtein radius at which the non-linear 
interactions of the brane bending mode become important in a 
cosmological background. 
The Vainshtein radius in the early universe is much smaller than 
the one in the Minkowski background, but in a self-accelerating 
universe it is the same as the Minkowski background.  
Our result shows that the perturbative approach is applicable 
beyond the Vainshtein radius for weak gravity by taking 
into account the second order effects of the brane bending mode.
The linearised cosmological perturbations are shown to be smoothly matched 
to the solutions inside the Vainshtein radius. 
We emphasize the importance of imposing a regularity condition 
in the bulk by solving the 5D perturbations and we highlight 
the problem of ad hoc assumptions on the bulk gravity that lead to 
different conclusions. 
\end{abstract}



\maketitle
\section{Introduction}
The acceleration of the late-time universe is one of 
the most important problems in cosmology. 
Within the framework of general relativity, the acceleration 
is supposed to be caused by unknown 
dark energy. The simplest option for dark energy is 
vacuum energy, but it is hard to explain why the 
vacuum energy is so small compared with the prediction
of particle physics. An alternative to dark energy is provided by 
models where large-distance modifications of gravity explain the 
acceleration. Probably the most widely studied example 
of a modified gravity model is the Dvali-Gabadadze-Porrati
(DGP) brane-world model in which gravity leaks off the 
4D brane into the 5D bulk spacetime \cite{DGP}. The 5D action describing 
the DGP model is given by
\be
S = \frac{1}{2 \kappa^2}\int d^5 x \sqrt{-g} {}^{(5)\!} R 
+ \frac{1}{2 \kappa_4^2} \int d^4 x \sqrt{-\gamma} R 
+ \int d^4 x \sqrt{-\gamma} {\cal L}_m
+ \frac{1}{\kappa^2} 
\int d^4 x \sqrt{-\gamma}  K,
\ee
where ${\cal L}_m$ is the Lagrangian for matter on the brane, 
$K_{\mu \nu}$ is the extrinsic curvature and $K=K^{\mu}_{\:\: \mu}$. 
The transition from 4D gravity to 5D gravity is governed by a crossover 
scale $r_c$, 
\be
r_c = \frac{\kappa^2}{2 \kappa_4^2},
\ee
which is the only parameter in this model.
A striking feature of this model is the existence of a
solution where the acceleration of universe is caused entirely
by gravity without introducing the cosmological constant \cite{cosmology}. 
In this solution the Hubble parameter approaches a
constant, $H \to 1/r_c$, at late times, mimicking the cosmological constant. 
This self-accelerating solution has attracted significant 
interest recently \cite{DGPp}. 

Unfortunately, it has been shown that the self-accelerating 
universe contains a ghost \cite{effective1, effective2, Koyama1, 
Koyama2, Charmousis, Izumi}. 
The existence of the ghost was shown rigorously on a de Sitter spacetime 
by studying linearised gravity.
Recently, however, there are some claims that the 
non-linear interactions obscure the conclusion 
on the existence of the ghost \cite{Dvali, DGI}. 
It has been recognized that the non-liner interactions 
of gravity in this model are much more subtle than 
4D general relativity \cite{nonlinear, Lue, Gruzinov, LS1, LS2, Rubakov, 
Tanaka, MS, domain, effective1, effective2}. 
The reason is that the graviton contains a scalar degree of freedom and the non-linear 
interaction of this mode becomes important on much larger scales than 
the usual graviton. This is analogous to the massive gravity 
model, where a helicity-0 mode becomes strongly coupled 
on very large scales for small graviton mass \cite{massive}. 
In the DGP model, the scalar mode
is a mix of the helicity-0 mode of the spin-2 5D graviton and 
the spin-0 mode called the radion \cite{Koyama1, Koyama2}. 
Physically, the scalar mode describes the bending of the brane in the bulk 
\cite{effective1, effective2, Tanaka}. 
It was shown that the non-linear interaction of the
brane bending mode becomes important at the so-called 
Vainshtein radius $r_*= (r_g r_c^2)^{1/3}$
where $r_g$ is the Schwarzschild radius of the source \cite{nonlinear}.
If we want to explain the late-time acceleration, 
we should require $r_c \sim H_0^{-1}$. 

One argument against the 
validity of the linearised analysis is that, 
for cosmology, $r_g$ is roughly the Hubble scale today 
$r_g \sim H_0^{-1}$, then the Vainshtein radius is also the 
horizon scale $r_* \sim H_0^{-1}$, which may indicate that the 
linearised cosmological perturbations are not 
valid \cite{Dvali}. However, most of the literature so far studied perturbations 
around Minkowski spacetime. It is still unclear 
what is the Vainshtein radius in a {\it cosmological background}. 
This is an important question to be addressed because 
the ghost exists in the self-accelerating solution where the 
Minkowski spacetime is not a solution. 
There is no ghost in a Minkowski brane in the DGP model.
Thus it is important 
to study non-linear interactions of the brane bending mode 
in a Friedmann background.

There is also a claim that the perturbative approach cannot 
be applied in the DGP model \cite{DGI}. This argument is based on 
the Schwarzschild solution obtained in Ref.~\cite{GI2}, which  
does not recover the linearised solution in the region $r>r_*$. 
However, this solution is obtained by closing the equations
on the brane by imposing ad hoc assumptions on the bulk gravity.  
In Ref.~\cite{KM}, it was shown that it is crucial to impose a proper 
boundary condition in the bulk to determine the behaviour 
of gravity on the brane. 

The aim of this paper is to study the non-linear interactions 
of the brane bending mode on a Friedmann background. We build on 
Ref.~\cite{KM} where linearised perturbations are solved properly 
in 5D spacetime. We extend the analysis of Ref.~\cite{KM}
by taking into account the second order effects of the brane 
bending mode. Then we study whether the linearized cosmological
perturbations can be smoothly matched to the solutions inside 
the Vainshtein radius. It should be noted that 
non-linear interactions on a Friedmann background 
were studied in Ref.~\cite{LS2} assuming spherical symmetry
and the modified Vainshtein radius was identified. We will 
confirm their result by properly solving the 5D metric 
perturbations without closing the equations on the brane
in an heuristic way in the same spirit as Ref.~\cite{KM}. 
For this purpose we closely follow the approach of Ref.~\cite{MS}, which 
studied weak gravity on the Minkowski background. 

\section{Quasi-static perturbations}
\subsection{Basic equations}
In this paper, we focus on weak gravity sourced by 
quasi-static matter fluctuations in a cosmological 
background. This analysis can be applied to describe 
the metric sufficiently far from a local source located 
in a cosmological background. We can also study the 
cosmological perturbations on sub-horizon 
scales in the matter-dominated era, which are relevant for the 
structure formation problem. 

The first order metric in the bulk is given in a 5D longitudinal gauge by
\begin{equation}
ds^2 = - (1 + 2 {\cal A}) N(t,y)^2 dt^2 + (1+2 {\cal R}) A(t,y)^2 
\delta_{ij} dx^{i} dx^{j} 
+ (1 +2 A_{yy}) dy^2,
\end{equation}
where
\begin{equation}
A(y,t)= a(t) (1 \mp Hy), \quad 
N(y,t)= 1 \mp H \left(1+\frac{\dot{H}}{H^2} \right)y,
\end{equation}
are the solutions for the background metric \cite{cosmology}. Note that 
the $(y,t)$-component of the metric can be neglected for a static 
source.
The Hubble parameter is determined by the Friedmann equation 
and the continuity equation:
\begin{equation}
\mp \frac{H}{r_c} = H^2 - \frac{\kappa_4^2}{3} \rho, 
\quad \dot{\rho} + 3 H(\rho+p) \rho =0.
\end{equation}
The solution with $-$ sign is called the normal branch solution
and the solution with $+$ sign is called the self-accelerating 
solution because there is a de Sitter solution even without any 
kind of matter on the brane \cite{cosmology}. The self-accelerating 
solution attracted significant interest as a model for 
dark energy from the large-distance modification 
of gravity.    

In the 5D Longitudinal gauge, the brane is not located at 
$y=0$ \cite{deffayet}. Then it is more convenient to move to a gauge where 
the brane is located at $y=0$. We perform a gauge 
transformation $y \to y - r_c \xi$, where $\xi$
is a scalar function describing the perturbation of the 
brane location, which is often called the brane bending mode. 
The resultant metric has the form
\begin{equation}
ds^2 = - (1 + 2 \Psi) N(t,y)^2 dt^2 + (1+2 \Phi) A(t,y)^2 
\delta_{ij} dx^{i} dx^{j} + 2 r_c \varphi_{,i} dx^i dy 
+ (1 +2 \Gamma) dy^2.
\end{equation}
At first order, $\varphi$ is identified as the brane bending mode $\xi$.
We are interested in perturbations well 
inside the horizon. Thus we will neglect all subleading terms 
suppressed by $aH/k \ll 1$, where $k$ is the 3D wavenumber 
of the perturbations. Within quasi-static approximations, 
time-derivative terms can be neglected.
We also neglect terms like $(A'/A) \Phi'$ where 
prime denotes a derivative with respect to $y$.
This is based on an assumption that $\Phi' \sim
k \Phi$. This assumption will be justified later. 
Although we are dealing with the linearised metric 
perturbations, it has been recognized that 
second order terms of $\varphi$ can be important 
on larger scales compared with the other second order 
contributions \cite{effective1, effective2, MS, Tanaka}. 
Thus we only keep the second order terms for $\varphi$. This assumption 
will also be verified later.

Under these assumptions, the 5D Einstein equations are given by:
\begin{eqnarray}
\delta {}^{(5)} G^{y}_{y} &=&\frac{1}{A^2} \nabla^2 \Psi 
+ \frac{1}{A^2} \nabla^2 \Phi - \frac{r_c}{A^2}
\left(2 \frac{A'}{A} +\frac{N'}{N} \right)
\nabla^2 \varphi
+\frac{r_c^2}{2 A^4} \left[    
(\nabla^2 \varphi)^2 -(\nabla_i \nabla_j \varphi)^2
\right]=0,
\label{eq:gyy} \\
\delta {}^{(5)} G^y_{i}&=&-(\Psi' + 2 \Phi')_{,i} -\frac{r_c^2}{2A^4} \left [  
(\nabla^j \varphi)(\nabla_j \nabla_i \varphi')
-(\nabla_i \varphi)(\nabla^2 \varphi')
\right]=0,
\label{eq:gyi} \\
\delta {}^{(5)}  G^t_t &=& 3 \Phi''  + \frac{2}{A^2} \nabla^2 \Phi +
\frac{\nabla^2}{A^2} (\Gamma -r_c \varphi') 
-2 \frac{r_c}{A^2} \left(\frac{A'}{A} \right)
\nabla^2 \varphi
+ \frac{r_c^2}{2 A^4} \left[ (\nabla^2 \varphi)^2 -(\nabla_i \nabla_j \varphi)^2
\right]=0, 
\label{eq:gtt} \\
\delta {}^{(5)} G^{i}_{j} &=& -\frac{1}{A^2}
(\nabla^i \nabla_j -\delta^i_j \nabla^2)
(\Phi+\Psi+\Gamma -r_c \varphi') + \delta^i_j
(\Psi'' + 2 \Phi'') 
+ \frac{r_c}{A^2}(\nabla^i\nabla_j - \delta^i_j 
\nabla^2 ) \left(\frac{A'}{A}+\frac{N'}{N} \right)
\varphi\\
&&
- \frac{r_c^2}{A^4} \left[     
(\nabla^2 \varphi)(\nabla^i \nabla_j \varphi) -(\nabla_j \nabla^k 
\varphi)(\nabla^i \nabla_k \varphi) 
\right]
+ \frac{1}{2} \delta^i_j \frac{r_c^2}{A^4} 
\left[(\nabla^2 \varphi)^2 -(\nabla_k \nabla_l \varphi)^2 
\right]=0.
\label{eq:gij}
\end{eqnarray}
For the spatial components $\delta {}^{(5)}G^i_j$, the trace of the 
equation gives 
\begin{equation}
\frac{2}{A^2} \nabla^2 (\Phi + \Psi + \Gamma - r_c \varphi')
- \frac{2 r_c}{A^2} \left(\frac{A'}{A}+\frac{N'}{N} \right)
\nabla^2 \varphi 
+ 3 (\Psi'' + 2 \Phi'') + 
\frac{1}{2} \frac{r_c^2}{A^4}\left[      
(\nabla^2 \varphi)^2 -(\nabla_i \nabla_j \varphi)^2
\right]=0.
\label{eq:trace}
\end{equation}
On the other hand, taking the divergence of the 
traceless part of $\delta {}^{(5)} G^{i}_j$, we get
 \begin{equation}
\frac{\nabla^2}{A^2}
(\Phi + \Psi + \Gamma -r_c \varphi')- \frac{r_c}{A^2}
\left(\frac{A'}{A}+\frac{N'}{N} \right) \nabla^2 \varphi
+ \frac{1}{4 }\frac{r_c^2}{A^4}
\left[(\nabla^2 \varphi)^2 -(\nabla_i \nabla_j \varphi)^2
\right]=0.
\label{eq:traceless}
\end{equation}

The existence of the brane imposes the junction condition 
at the brane, that relates the extrinsic curvature
with the energy-momentum tensor on the brane
\be
K_{\mu \nu} - K g_{\mu \nu} = -\frac{\kappa^2}{2} T_{\mu \nu} + r_c G_{\mu \nu}.
\label{junction}
\ee
We should note that due to the induced gravity term, the Einstein tensor 
appears in the junction condition. 
The $(t,t)$ component of the junction condition 
Eq.~(\ref{junction}) gives 
\begin{equation}
\frac{2}{a^2} \nabla^2 \Phi 
= - \kappa_4^2 \delta \rho 
+ \frac{1}{a^2} \nabla^2\varphi 
- \frac{3}{r_c} \Phi'.
\label{jun:tt} 
\end{equation}
The spatial components give
\begin{eqnarray}
\Phi + \Psi &=& \varphi, \\
\label{jun:traceless}
\Psi' + 2 \Phi' &=& 0.
\label{jun:trace}
\end{eqnarray}

\subsection{Solutions in the bulk}
Let us first solve the perturbations in the bulk.
Combining Eq.~(\ref{eq:gyy}) and Eq.~(\ref{eq:traceless}), 
$\Phi$ and $\Gamma-r_c \varphi'$ are written in terms of $\Psi$ 
and $\varphi$:
\begin{eqnarray}
\frac{\nabla^2}{A^2} \Phi
 &=& -\frac{1}{2} \frac{\nabla^2}{A^2} \Psi
 + \frac{r_c}{2 A^2} \left(2 \frac{A'}{A} + \frac{N'}{N} 
 \right) \nabla^2 \varphi 
 - \frac{r_c^2}{4 A^4} 
 \left[     
 (\nabla^2 \varphi)^2 - (\nabla_i \nabla_j \varphi)^2
 \right], \label{Phi} \\
\frac{\nabla^2}{A^2} (\Gamma -r_c \varphi') 
&=& -\frac{1}{2} \frac{\nabla^2}{A^2} \Psi
+ \frac{r_c}{2 A^2} \left(\frac{N'}{N} \right)\nabla^2 \varphi.
\label{G}
\end{eqnarray}
Consistency between Eqs.~(\ref{eq:gyi}) and (\ref{Phi}) 
requires 
\begin{equation}
\varphi' =0, \quad \Psi' + 2 \Phi' =0.
\label{varphi}
\end{equation}
The latter is consistent with the junction condition Eq.~(\ref{jun:trace}).
Then substituting Eqs~(\ref{Phi}) and (\ref{G}) into Eq.~(\ref{eq:gtt})
and using Eq.~(\ref{varphi}), we get a wave equation for $\Psi$
\begin{equation}
\Psi'' + \frac{\nabla^2}{A^2} \Psi - 
\left(\frac{N'}{N} \right) \frac{r_c}{A^2} \nabla^2 \varphi =0.
\label{wave}
\end{equation}
By performing a Fourier transformation, the solution is 
given  by
\begin{equation}
{\cal A} =\Psi - \frac{N'}{N} r_c \varphi
= \left[ c_1  (1 \mp Hy)^{\pm k/aH} + c_2 (1 \mp Hy)^{\mp k/aH} \right],
\label{sol}
\end{equation}
for a given $k$, with our approximation $k/aH \gg 1$. We impose the regularity 
condition in the bulk so that the perturbations do not diverge
at $y \to \infty$ in the self-accelerating branch, and 
$y = 1/H$ in the normal branch. This means that we take 
$c_2 =0$ \cite{KM}. We should note that the regularity condition 
verifies our assumption that the terms like $(A'/A) \Phi'$ 
can be neglected compared with the terms like $\nabla^2 \Phi/A^2$,  
with our approximation $k/aH \gg 1$.

\subsection{Equations on the brane}
Now we impose the junction conditions. From Eqs.~(\ref{sol})
and (\ref{Phi}), it is possible to show that 
\begin{equation}
\frac{\Phi'}{r_c} \sim \frac{k}{a r_c} \Phi 
\ll \frac{k^2}{a^2} \Phi,
\end{equation}
for perturbations whose physical wavelengths are shorter than 
$r_c$, $k r_c/a\gg 1$. Thus we can neglect $\Phi'$ in 
the junction condition Eq.~(\ref{jun:tt}). Then the 
projection of Eq.~(\ref{eq:gyy}) on the brane and the 
junction conditions Eqs.~(\ref{jun:tt}) and (\ref{jun:traceless}),
provide a closed set of equations on the brane for 
$\Phi$, $\Psi$ and $\varphi$. The effective Einstein equations are
written as 
\begin{eqnarray}
\frac{2}{a^2} \nabla^2 \Phi &=& -\kappa_4^2 \delta \rho 
+ \frac{1}{a^2} \nabla^2 \varphi, 
\label{einstein:tt} \\
\Psi +\Phi &=& \varphi,
\label{einstein:traceless} 
\end{eqnarray}
and the equation of motion for $\varphi$ is given by
\begin{equation}
3 \beta(t)
\frac{\nabla^2}{a^2} \varphi 
+ \frac{r_c^2}{a^4} 
\left[      
(\nabla^2 \varphi)^2 - (\nabla_i \nabla_j \varphi)^2 
\right]=\kappa_4^2 \delta \rho,
\label{eq:phi}
\end{equation}
where 
\begin{equation}
\beta(t) =1-\frac{2r_c}{3}\left(     
2 \frac{A'}{A} + \frac{N'}{N} \right)
= 1 \pm 2 Hr_c \left( 1 + \frac{\dot{H}}{3 H^2} \right).
\label{beta}
\end{equation}
Here the $+$ sign corresponds to the normal branch and
 the $-$ sign to the self-accelerating one.

\section{Solutions on a brane}
\subsection{Linearised solutions}
We begin with linearised solutions by  
neglecting the second order contributions of $\varphi$.
The solutions for the metric perturbations are easily 
obtained as 
\begin{eqnarray}
\frac{\nabla^2}{a^2} \Phi &=& -\frac{\kappa_4^2}{2} 
\left(1 - \frac{1}{3 \beta} \right) \delta \rho, 
\label{solution:linear1}
\\
\frac{\nabla^2}{a^2} \Psi &=& \frac{\kappa_4^2}{2} 
\left(1 + \frac{1}{3 \beta} \right) \delta \rho, 
\label{solution:linear2}
\end{eqnarray}
which agree with the solutions obtained in Ref.~\cite{LS2,KM}.

The linearised equations can be described by a Brans-Dicke
(BD) theory. The perturbed Einstein equations in the BD theory 
are given by
\begin{equation}
\delta G_{\mu \nu} = - (\nabla_{\mu} \nabla_{\nu} - g_{\mu \nu} \nabla^2)
\varphi,
\label{BD1}
\end{equation}
and the equation of motion for the BD scalar is 
\begin{equation}
\frac{\nabla^2}{a^2} \varphi = \frac{\kappa_4^2}{3 +2 \omega} \delta \rho,
\label{BD2}
\end{equation}
where $\omega$ is the BD parameter.
Comparing Eqs.~(\ref{einstein:tt}) - (\ref{eq:phi}) with Eqs.(\ref{BD1}) and (\ref{BD2}), 
we find that the brane bending mode acts as the BD scalar 
and the BD parameter is given by \cite{LS2,KM,KMi}
\begin{equation}
\omega = \frac{3}{2} (\beta -1).
\end{equation}

The sign of $\beta$ is directly related to the 
existence of the ghost in de Sitter spacetime.
In the self-accelerating branch, $\beta$ is negative
for $Hr_c >1/2$. In the BD theory, the BD scalar has 
the wrong sign for the kinetic term if $\omega < -3/2$,
that is $\beta <0$.  The condition that $\beta$ is negative 
is given by $Hr_c >1/2$, which is precisely the condition
for the existence of the ghost in de Sitter spacetime, as was shown 
in Refs~\cite{effective1, effective2, Koyama1}. 
On the other hand, in the normal branch, $\beta$ is 
positive and we expect no ghost in this branch of the solutions.

\subsection{Spherically symmetric solutions}
Next, we study the effect of the second-order 
contributions of $\varphi$. Unfortunately, 
it is not easy to solve the equations for $\varphi$ 
with the non-linear interactions. Thus we assume 
spherical symmetry to simplify the problem.
The equation for $\varphi$ (\ref{eq:phi}) is then given by 
\begin{equation}
\left(\frac{d^2}{d r^2} + \frac{2}{r} \frac{d}{dr}
\right) (3 \beta \varphi + \Xi) = \kappa_4^2 \delta \rho,
\end{equation}
where
\begin{equation}
\Xi = 2 r_c^2 \int \frac{1}{r} 
\left( \frac{d \varphi}{d r} \right)^2 dr,
\end{equation}
in agreement with Ref.~\cite{MS} in a Minkowski spacetime.
Let us consider a source localized in some compact region. 
Then it is possible to integrate the equation
to get 
\begin{equation}
3 \beta \varphi + \Xi + \frac{r_g}{r}=0,
\label{phiinteg} 
\end{equation}
where 
\begin{equation}
r_g = \kappa_4^2 \int^r_0 dr r^2 \delta \rho,
\end{equation}
is the Schwarzschild radius of the source. 
Hereafter, we assume $r_g=$ const, for simplicity. 
Taking the $r$ derivative of Eq.~(\ref{phiinteg}) gives an algebraic 
equation for $d \varphi/dr$. Then we get a solution for 
$d \varphi/dr$ as 
\begin{equation}
\frac{d\varphi}{dr} = \frac{r_g}{r^2} \Delta(r), 
\quad \Delta(r) = \frac{2}{3 \beta} \left( \frac{r}{r_*} \right)^3 
\left(\sqrt{1+ \left(\frac{r_*}{r}\right)^3} -1 \right),
\label{spherical:phi}
\end{equation}
where
\begin{equation}
r_*= \left(\frac{8 r_c^2 r_g}{9 \beta^2} \right)^{1/3},
\end{equation}
which is the Vainshtein radius for a source in a cosmological background.
This is in agreement with the result of Ref.~\cite{LS2}, but 
we arrive at this result by solving the 5D 
bulk metric and imposing the regularity condition in the bulk,
without closing the equations on the brane in an heuristic 
way. The solutions for the metric perturbations can be obtained as 
\begin{eqnarray}
\Phi &=& \frac{r_g}{2r} + \frac{\varphi}{2},\\
\Psi &=& -\frac{r_g}{2r} + \frac{\varphi}{2}.
\label{spherical:metric}
\end{eqnarray}

On scales larger than the Vainshtein radius 
$r > r_*$, the solutions are given by
\begin{eqnarray}
\Phi &=& \frac{r_g}{2r} 
\left(1- \frac{1}{3 \beta} \right), \\ 
\Psi &=& -\frac{r_g}{2r}
\left(1+ \frac{1}{3 \beta} \right).
\end{eqnarray}
which agree with the linearised solutions Eqs.~(\ref{solution:linear1})
and (\ref{solution:linear2}). 
This shows that the linearised solutions do make sense as long as we are 
considering scales larger than the Vainshtein radius. 

On scales smaller than the Vainshtein radius, $r < r_*$, the solutions for 
$\Psi$ and $\Phi$ are obtained as 
\begin{eqnarray}
\Phi &=& \frac{r_g}{2r} + \frac{1}{\beta} 
\sqrt{\frac{\beta^2 R_g r}{2r_c^2}}, \\
\Psi &=& -\frac{r_g}{2r} + \frac{1}{\beta} 
\sqrt{\frac{\beta^2 R_g r}{2 r_c^2}}.
\label{einsteinphase}
\end{eqnarray}
In this region, the corrections to the solution in 4D general 
relativity are suppressed for $r < r_*$ so that Einstein gravity 
is recovered. From Eq.~(\ref{eq:phi}), we can see that $\Xi$ dominates 
over the linear term in this region. This indicates 
that once $\varphi$ becomes non-linear, the solutions for 
the metric approach those in 4D general relativity. 
We should note that $\beta$ is negative in the self-accelerating 
solution while $\beta$ is positive in the normal branch solution.
Then the corrections to 4D general relativity solutions
have opposite signs in these solutions, as was first pointed 
out in Ref~\cite{LS1}. By a simple coordinate transformation,
we can check that our solutions agree with the results of 
Ref.~\cite{LS2}. 

Even on scales smaller than the Vainshtein radius $r < r_*$, 
the induced metric perturbations are small as long as we consider 
scales larger than the Schwartzschild radius $r > r_g$. This justifies 
our assumption of neglecting all second order contributions 
other than the second order terms of $\varphi$. It should be 
also emphasized that the $(y,r)$ component of the metric, 
$r_c \varphi_{,r}$, is evaluated as 
\begin{eqnarray}
(r_c \varphi_{,r})^2 &\sim& \left( \frac{r_*}{r} \right)^3 \left(    
\frac{r_g}{r} \right), \quad \mbox{for} \;\;\; r > r_*,\\
(r_c \varphi_{,r})^2 &\sim& \left(\frac{r_g}{r} \right), 
\quad \mbox{for} \;\;\; r < r_*.
\label{phir}
\end{eqnarray}
The higher order terms of $\varphi$ in the Einstein equations have 
higher order powers of $r_c \varphi_{,r}$. Thus they are suppressed 
for $r > r_g$. Then we only need to keep the second order terms 
which can be comparable to the linear terms as is seen from 
Eqs.~(\ref{einsteinphase}) and (\ref{phir}).

\subsection{Cosmological perturbations}
Finally, we consider cosmological perturbations in a matter-dominated 
universe. We define an over-density of dark matter as 
\begin{equation}
\delta = \frac{\delta \rho}{\rho}.
\end{equation}
The continuity equation and the Euler equation are the same 
as in 4D general relativity:
\begin{eqnarray}
\frac{\partial \delta}{\partial t} + \frac{1}{a} \nabla(1+\delta) 
{\bf v}&=&0, 
\label{matter1}\\
\frac{\partial {\bf v}}{ \partial t} + \frac{1}{a}
({\bf v} \cdot \nabla) \cdot {\bf v} + H {\bf v} 
&=& - \frac{1}{a} \nabla \Psi,
\label{matter2}
\end{eqnarray}
where ${\bf v}$ is the velocity perturbation of dark matter.
Here we introduce time-derivative terms.
In order to ensure our quasi-static approximation, the 
time-dependence of the over-density $\delta$ 
should be weak, $\partial_t \delta \ll k \delta$,
which is indeed valid for dust matter. 
Combining Eqs.~(\ref{matter1}) and (\ref{matter2}) with 
Eqs.~(\ref{einstein:tt}), (\ref{einstein:traceless}) and 
(\ref{eq:phi}), we can describe the evolution of the 
dark matter over-density.  

The non-linear terms in the equation for $\varphi$, 
Eq.~(\ref{eq:phi}), become dominant when  
\begin{equation}
\frac{\beta a^2}{r_c^2 k^2} < \varphi.
\end{equation}
Using the linear term in Eq.~(\ref{eq:phi}), $\varphi$ is estimated as 
\begin{equation}
\varphi \sim \frac{H^2 a^2}{\beta k^2} \delta,
\end{equation}
where we used $\kappa_4^2 \rho \sim H^2$.
Then the condition that the non-linear terms become important 
is given in terms of $\delta$ by
\begin{equation}
\beta^2 (Hr_c)^{-2} \sim O(1) < \delta.
\end{equation}
If non-linear terms become dominant, $\varphi$ is estimated 
as 
\begin{equation}
\frac{k^2}{a^2} \varphi \sim \left(\frac{H^2 a^2}{ r_c^2 k^2}\right) \delta.
\end{equation}
Then in the Poisson equation Eq.~(\ref{einstein:tt}), the contribution of 
$\varphi$ can be neglected and 4D general relativity is recovered. 
Thus, from these rough estimations, we expect to recover 
4D general relativity for non-linear over-density $\delta \gg 1$. 
This also means that for linear over-density $\delta \ll 1$, the 
second order terms of $\varphi$ can be neglected and the 
linearised cosmological perturbations do perfectly make sense
as opposed to the claim made in Ref.~\cite{Dvali}. In order to verify these 
estimations, one should solve the non-linear equations for 
$\delta$ and $\varphi$, which is difficult even in conventional 
4D general relativity. One approach is to consider the spherically 
symmetric collapse of the over-density. This was done in 
Ref.~\cite{LS2}, and it was demonstrated that once the over-density 
exceeds $O(1)$, 4D general relativity is recovered. This confirms
our estimations. 

\section{Effective theory on the brane}
\subsection{Effective theory for $\varphi$}
In the previous section, we find that the brane bending 
mode $\varphi$ plays a crucial role in the DGP model. 
It is possible to understand the role of the brane bending mode 
in a covariant way as was shown in Refs.~\cite{effective1, effective2}. 
We begin with the definition of the extrinsic curvature: 
\begin{equation}
K_{\mu \nu} = \frac{1}{2 {\cal N}} 
( \partial_y g_{\mu \nu} - \nabla_{\mu} N_{\nu} -
\nabla_{\nu} N_{\mu} ),
\end{equation}
where $g_{\mu \nu}$ is the induced metric, $N_{\mu}$ is 
a shift function and ${\cal N} = \sqrt{g_{yy} - N_{\mu} N^{\mu}}$
is a lapse function. 
Let us first consider perturbations around Minkowski 
spacetime
\begin{equation}
g_{\mu \nu} = \eta_{\mu \nu} + \delta g_{\mu \nu},
\end{equation}
The lapse function is given in terms of the 
brane bending mode $\varphi$ by $N_{\mu} = r_c \nabla_{\mu} \varphi$.
Then the extrinsic curvature is given in term of $\varphi$ by
\begin{equation}
\delta K_{\mu \nu} = -r_c \nabla_{\mu} \nabla_{\nu} \varphi.
\label{kphi}
\end{equation}

An important result obtained by solving the 5D perturbations
is that we can neglect the $y$ derivative of the induced metric 
in the junction condition because
\begin{equation}
\partial_y g_{\mu \nu} \ll r_c \nabla^2 \delta g_{\mu \nu}.
\end{equation}
Then the junction condition becomes
\begin{equation}
\delta G_{\mu \nu} = \kappa_4^2 \delta T_{\mu \nu} 
- (\nabla_{\mu} \nabla_{\nu} - g_{\mu \nu} \nabla^2 ) 
\varphi. 
\label{ein:phi}
\end{equation}
On the other hand, the Gauss equation in the bulk, that is 
the $(y,y)$ component of the 5D Einstein equations, gives 
\begin{equation}
R - K^2 + K_{\mu \nu} K^{\mu \nu} =0.
\label{gauss}
\end{equation}
Then combining  Eqs.~(\ref{ein:phi}) and (\ref{gauss}), we get 
the equation for $\varphi$ as 
\begin{equation}
3 \nabla^2 \varphi + r_c^2 \left[     
(\nabla^2 \varphi)^2 -(\nabla_{\mu} \nabla_{\nu} \varphi)^2
\right] =-\kappa_4^2 T,
\end{equation}
which reproduces Eq.~(\ref{eq:phi}) for static perturbations.  
We should emphasize that the non-linear terms for $\varphi$ 
come from the non-linear terms of $K_{\mu \nu}$. Even if 
we are dealing with weak gravity where the induced 
curvature is small, this does not necessarily mean that 
the non-linearity of $K_{\mu \nu}$ can be neglected. 
We should also note that the higher order terms in $\varphi$
comes from $N_{\mu} N^{\mu}$ which is given by 
$(r_c \varphi_{,r})^2$ in a spherically symmetric spacetime. 
We have shown that these contributions are suppressed 
as long as $r > r_g$. 

In a cosmological background, the extrinsic curvature 
has contributions from the background
\begin{equation}
K_{tt} = - \frac{N'}{N} \quad 
K_{ij} = \frac{A'}{A} \delta_{ij}.
\label{kback}
\end{equation}
This gives an additional first order contribution in the 
Gauss equation \cite{effective2, deffayet2}
\begin{equation}
\delta (- K^2 + K_{\mu \nu}K^{\mu \nu})
= 2r_c \left(\frac{N'}{N} + 2 \frac{A'}{A}\right)
\nabla^2 \varphi.
\end{equation}
This modifies the coefficient of the linear kinetic term for $\varphi$ 
to $3 \beta$. In de Sitter spacetime, this is exactly the origin of 
the ghost in the self-accelerating solution. 

The equation of motion for $\varphi$ can be derived from the action
\begin{equation}
S \propto  -\int d^4 x \sqrt{-\gamma}
\Big[ 
3 \beta (\nabla \varphi)^2 + r_c^2 (\nabla \varphi)^2 \nabla^2 \varphi
\Big],
\end{equation}
assuming static perturbations.
Defining a new field $\pi$ as $\pi = M_4 \varphi$, where $\kappa_4^2 
=1/M_4^2$, the action can be rewritten as 
\begin{equation}
S \propto -\int d^4 x \sqrt{-\gamma}
\Big[ 
3 \beta (\nabla \pi)^2 + \frac{1}{\Lambda^3} (\nabla \pi)^2 \nabla^2 \pi
\Big],
\label{action}
\end{equation}
where $\Lambda = (M_4/r_c^2)^{1/3}$. In de Sitter spacetime, this agrees 
with the boundary effective action for the brane bending mode derived 
in Ref.~\cite{effective1, effective2}. Thus our solution is consistent with the 
effective theory for the brane bending mode of Refs.~\cite{effective1, effective2}.

\subsection{Effective equation on the brane}
It is also possible to construct an effective theory for $\varphi$ 
using the effective equations on the brane. 
Projecting the 5D Einstein equations on the brane, 
the effective equations are given by \cite{SMS}
\begin{equation}
G_{\mu\nu}= \kappa^4 \Pi_{\mu\nu} -E_{\mu\nu},
\label{effective}
\end{equation}
where 
\begin{equation}
\Pi_{\mu\nu} = -\frac{1}{4} \tilde{T}_{\mu \alpha} \tilde{T}
_{\nu}^{\alpha} + \frac{1}{12} \tilde{T} \tilde{T}_{\mu \nu}
+\frac{1}{8} g_{\mu \nu} \tilde{T}_{\alpha \beta} \tilde{T}^{\alpha \beta}
-\frac{1}{24} g_{\mu \nu} \tilde{T}^2,
\end{equation}
\begin{equation}
\tilde T_{\mu \nu} = T_{\mu \nu} - \kappa_4^{-2} G_{\mu \nu},
\end{equation}
and $E_{\mu \nu}$ is the projection of electric part of the 
5D Weyl tensor. 
For fluctuations around the vacuum Minkowski spacetime, 
$G_{\mu \nu}$ is written solely in terms of 
$\varphi$ from Eq.~(\ref{ein:phi}). Thus the effective equations are written 
in terms of $\varphi$ except for $E_{\mu \nu}$.
The resultant effective equations are \cite{GI1}
\begin{equation}
-(\nabla_{\mu} \nabla_{\nu} - g_{\mu \nu} \nabla^2)\varphi
= -\frac{r_c^2}{2} 
\Big[    
g_{\mu \nu} \left\{(\nabla^2 \varphi)^2 - (\nabla_{\alpha} \nabla_{\beta} \varphi)^2 \right\} -2 \left\{(\nabla^2 \varphi)(\nabla_{\mu} \nabla_{\nu}
\varphi) -(\nabla_{\mu} \nabla_{\alpha} \varphi)(\nabla_{\nu} 
\nabla^{\alpha} \varphi )  \right\} \Big] -E_{\mu \nu}.
\end{equation}
Taking the trace of this equation gives 
\begin{equation}
3 \nabla^2 \varphi + r_c^2 
\left[(\nabla^2 \varphi)^2 - (\nabla_{\mu} \nabla_{\nu} \varphi)^2 
\right] =0,
\label{trace}
\end{equation}
because $E_{\mu \nu}$ is traceless. This reproduces the equation 
of motion for $\varphi$, Eq.~(\ref{eq:phi}). 
On the other hand, the $(t,t)$ component gives 
\begin{equation}
3 \nabla^2 \varphi + \frac{3}{2} r_c^2 
\left[(\nabla^2 \varphi)^2 - (\nabla_{\mu} \nabla_{\nu} \varphi)^2 
\right] = 3 E_{tt}.
\end{equation}
If we neglected $E_{tt}$, this equation would contradict 
Eq.~(\ref{trace}) as is pointed out by Ref.~\cite{GI1}. 

However, we should {\it not} neglect $E_{\mu \nu}$. From 
the 5D metric, $E_{tt}$ is calculated as 
\begin{equation}
E_{tt} = -3 \Phi'' -\nabla^2 (\Gamma- r_c \varphi').
\end{equation}
It should be emphasized that $E_{tt}$ contains the second 
derivative of the metric with respect to $y$. Therefore, 
unless we solve the 5D perturbations, it is impossible 
to evaluate this term on the brane as the junction condition
on the brane only determines the first derivatives. 
Using the solutions Eqs.~(\ref{Phi}) and (\ref{G}) and 
the equations on the brane Eqs.~(\ref{einstein:tt}), 
(\ref{einstein:traceless}) and (\ref{eq:phi}), $E_{tt}$
is evaluated as 
\begin{equation}
E_{tt} = \frac{r_c^2}{6} \left[   
(\nabla^2 \varphi)^2 - (\nabla_{\mu} \nabla_{\nu} \varphi)^2
\right].
\end{equation}
Then it turns out that the $(t,t)$ component is fully consistent with 
Eq.~(\ref{trace}). Thus the effective equations (\ref{effective}) are 
consistent with our solutions. This is in fact trivial as the 
effective equations are nothing but the projection of 5D 
Einstein equations. Thus as long as we solve the 5D equations,
the solutions should trivially satisfy the effective equations.

\subsection{Condition on $E_{\mu \nu}$}
At linearised level, it was shown that the regularity 
condition for bulk perturbations gives a condition on 
$E_{\mu \nu}$ which cannot be determined by equations on 
the brane \cite{KM}. Here we check that this condition is not modified by the 
inclusion of non-linear interactions of $\varphi$.
First let us parameterize $E_{\mu \nu}$ as 
\begin{equation}
\delta E^t_t = \kappa_4^2 \delta \rho_E, \quad 
\delta E^i_j = -\kappa_4^2 \Big[ \frac{1}{3} \delta \rho_E \delta^{i}_j+ 
\delta \pi^{i}_{E \; j} \Big], 
\end{equation}
where $\delta \pi^i_{E \; j} = \nabla^i \nabla_j \delta \pi_E 
- (1/3) \delta^i_j \nabla^2 \pi_E$. 
In a cosmological background, these quantities are given by
\begin{eqnarray}
\kappa_4^2 \delta \rho_E &=& 3 \Phi'' + \frac{\nabla^2}{a^2} 
(\Gamma - r_c \varphi'), \\
\kappa_4^2 \delta \pi_E &=& \Gamma - r_c \varphi'.
\end{eqnarray}
Then using Eqs.~(\ref{Phi}), (\ref{G}) and (\ref{wave}), 
it is possible to show that these satisfy the condition
\begin{equation}
\delta \rho_E + 2 \frac{\nabla^2}{a^2} \delta \pi_E =0.
\label{condition}
\end{equation}
Note that we have already used the regularity condition 
to assume $\Phi' \sim k \Phi$ and neglect terms suppressed by 
$aH/k \ll 1$. 
This is exactly the condition obtained in Ref.~\cite{KM}. 
As in the Minkowski case, we can evaluate $\delta \rho_E$
as 
\begin{equation}
\kappa_4^2 \delta \rho_E 
= -\frac{1}{6} \frac{\Big[ 1 \pm 2 Hr_c \left(1 + \frac{\dot{H}}{H^2} \right) \Big]}
{\Big[ 1 \pm 2 Hr_c \left(1 +\frac{\dot{H}}{3 H^2}  \right) \Big]}
\frac{r_c^2}{a^4} \left[    
(\nabla^2 \varphi)^2 - (\nabla_{i} \nabla_{j} \varphi)^2
\right]
+ \frac{2}{3} \frac{\Big[ 1 \pm 2 Hr_c \left(1 + \frac{\dot{H}}{2 H^2} \right) \Big]}
{\Big[ 1 \pm 2 Hr_c \left(1 +\frac{\dot{H}}{3 H^2}  \right) \Big]} 
\kappa_4^2 \delta \rho.
\end{equation}
This agrees with the result obtained in Ref.~\cite{KM} if we neglect the 
second order terms of $\varphi$. 
The $(t,t)$ component of the effective Einstein equation gives
\begin{equation}
\delta G^{t}_t = 2 \frac{A'}{A} \frac{r_c}{a^2} \nabla^2 \varphi
- \frac{r_c^2}{2 a^4} \left[      
(\nabla^2 \varphi)^2 -(\nabla_i \nabla_j \varphi)^2 
\right] - \kappa_4^2 \delta \rho_E,
\end{equation}
where we used the expressions for $\Pi_{\mu \nu}$ in terms 
of the extrinsic curvature
\begin{equation}
\kappa^4 \Pi^{\mu}_{\nu} = K K^{\mu}_{\nu} -K^{\mu \rho}
K_{\nu \rho} -\frac{1}{2} \delta^{\mu}_{\nu}
(K^2 - K_{\alpha \beta} K^{\alpha \beta}),
\end{equation}
and Eqs.~(\ref{kphi}) and (\ref{kback}).  Using the solution for 
$\delta \rho_E$, it is possible to check that this equation reduces 
to Eq.~(\ref{einstein:tt}) using Eq.~(\ref{eq:phi}).

\section{Conclusion and Discussions}
In this paper, we studied quasi-static perturbations in a 
cosmological background in the DGP brane world. Using 
Gaussian coordinates, we derive the solutions for weak 
gravity by taking into account the non-linear interactions 
of the brane bending mode. Solving the bulk metric perturbations
and imposing a regularity condition, we got a closed set 
of equations, Eqs~.(\ref{einstein:tt}), (\ref{einstein:traceless})
and (\ref{eq:phi}) on the brane. At linearised level the theory 
is described by a BD theory with the BD parameter given by 
$\omega = 3 (\beta-1)/2$, where $\beta$ is given by Eq.~(\ref{beta}).
We studied the effects of non-liner interactions of the 
brane bending mode assuming spherical symmetry. We find that 
the Vainshtein radius at which non-linear interactions of the 
bending mode become important is given by $r_*^3 = r_V^3/\beta^2$
where $r_V$ is the Vainshtein radius in the Minkowski background. 
In the early universe, $\beta^2 \gg 1$, so the Vainshtein radius is very 
small. Note that in this limit, we recover 4D general relativity 
even at linearised level, as the BD parameter becomes large \cite{deffayet2, KMi}. 
On the other hand, in the self-accelerating universe, 
$Hr_c =1$, $\beta^2=1$, so the Vainshtein radius is the same as 
in the Minkowski background. On scales smaller than the 
Vainshtein radius, $r<r_*$, the solution approaches 4D general 
relativity. Our solutions agree with the results of Ref.~\cite{LS1,LS2} 
in the Friedmann background, and the results of Refs.~\cite{Gruzinov, MS, Tanaka} 
in the Minkowski background. 

Our equations can be applied to cosmological 
perturbations on subhorizon scales in the matter-dominated era. 
Although the non-linear equations are difficult 
to solve in this case, we can estimate the scale at which the non-linear 
interactions of the brane bending mode become important. We found that 
once the dark matter over-density becomes non-linear, the 
non-linear terms of the bending mode also become important 
and the behaviour of metric perturbations approach to 4D general 
relativity. This result is in accord with the finding in 
Ref.~\cite{LS2} where a spherical symmetric collapse is studied in 
the self-accelerating background. Our result indicates that the linearised 
cosmological perturbations analysis does make sense in the same way as 
in the conventional 4D cosmology as opposed to a claim made in Ref.~\cite{Dvali}. 

We checked the consistency of our solutions with the effective 
equations on the brane. First, we checked that our solutions can be 
derived from the boundary effective theory for the bending mode 
derived in Refs.~\cite{effective1, effective2}. 
Following Ref.~\cite{GI1}, we also checked the 
consistency of our solutions with the effective equations 
on the brane derived by a projection of 5D Einstein equations.
A key quantity is the electric part of the bulk Weyl tensor
projected onto the brane $E_{\mu \nu}$. 
If we neglected this Weyl contribution,
the effective equations were inconsistent. Using the solutions 
in the bulk, we can evaluate $E_{\mu \nu}$
on the brane. We have shown that once the contribution from 
$E_{\mu \nu}$ is properly taken into account, the effective 
equations are fully consistent. Our analysis is consistent
with the boundary effective action Eq.~(\ref{action}) at least 
for static perturbations. It was pointed out that this 
effective action manifests superluminal propagation
if we consider time-dependent fluctuations around a spherically 
symmetric solution \cite{super}. It would be important to extend 
our 5D analysis to include time-dependent perturbations 
to check the validity of the boundary effective 
action with time-dependent perturbations, and understand 
the causality of the propagation in the 5D spacetime. 

Our conclusion is different from that of Ref.~\cite{DGI}. 
In Ref.~\cite{DGI}, it is argued that the linearised perturbations, which 
by themselves are valid at $r>r_*$ are not guaranteed to match 
to the solution inside $r<r_*$. This argument is based on 
the Schwarzschild solution obtained in Ref.~\cite{GI2}, which  
does not recover the linearised solution in the region $r>r_*$. 
However, the Schwarzschild solution in Ref.~\cite{GI2} is not derived 
by solving the bulk metric and imposing a 
proper boundary condition in the bulk. Instead 
they imposed a specific form of the metric and closed the equations 
on the brane.  This is in fact the same as imposing an ad hoc condition 
on $E_{\mu \nu}$. As we have shown in this paper, the condition on 
$E_{\mu \nu}$ has to be determined by solving the bulk 
metric and imposing an appropriate boundary condition 
in the bulk. For weak gravity that is valid for $r > r_g$,
the regularity condition in the bulk uniquely determines a condition for 
$E_{\mu \nu}$, Eq.~(\ref{condition}). 
The Schwarzschild solution found in Ref.~\cite{GI2} does not 
satisfy this condition in the weak gravity region. 
Thus their solution is unlikely to describe  
weak gravity sourced by a physical local source on a brane. 
On the other hand, it is still an open question what is a proper condition 
on $E_{\mu \nu}$ for strong gravity. An outstanding open question is to find 
a fully non-linear spherically symmetric solution that properly reproduces the 
solutions Eqs.~(\ref{spherical:phi}) and (\ref{spherical:metric}) for weak gravity.
We will come back to this issue in a separate publication.

  Finally, we comment on the ghost problem. Our analysis shows that the 
linearised analysis does make sense as long as we consider scales 
beyond the Vainshtein radius $r_*$ for a local source. Then on scales 
$r > r_*$ we find a ghost in the self-accelerating universe. 
Usually, we expect an instant instability of the spacetime in the 
presence of the ghost. Then the self-accelerating universe would 
not be a viable background for cosmology. However, in this case, it is not 
so obvious that the ghost leads to an instant instability of 
the spacetime classically \cite{SSH}, or even quantum mechanically \cite{IK}. 
Furthermore, non-linear interactions of the bending mode would become 
important if instabilities kick in. Further study is 
needed to understand the fate of this ghost.
On the other hand, the normal branch solution is free from the ghost. Although 
the solution itself cannot be an alternative to dark energy, it still
provides an interesting possibility to realize an expansion history 
of the universe which is equivalent to a dark energy 
model with an equation of state less than $-1$ \cite{phantom}. This model also
provides a concrete example for the large distance modification of 
gravity \cite{Charmousis}. Our equations (\ref{einstein:tt}), (\ref{einstein:traceless})
and (\ref{eq:phi}) are the basis for the study of structure formation tests 
in this model. 

\section*{Acknowledgments}
KK is supported by PPARC. FPS is supported by
``Funda\c{c}\~{a}o para a Ci\^{e}ncia e a Tecnologia (Portugal)",  
with the fellowship's reference number: SFRH/BD/27249/2006.
We would like to thank Roy Maartens for discussions and 
a careful reading of this manuscript.

\end{document}